\documentclass[twocolumn,prd,aps,amsmath,amsfonts,showpacs,preprintnumbers,nofootinbib,showkeys]{revtex4}
\usepackage{graphicx, color}
\usepackage{longtable}
\usepackage{bbold}
\usepackage{amsmath}
\usepackage{subfigure}

\def\e{\mathrm{e}}
\def\p{\partial}
\def\I{\mathrm{i}}

\def\F{\mathcal{F}}
\def\erf{\mathrm{erf}}

\def\nquad{\!\!}
\def\loOm{{\stackrel{}{t}}}

\newcommand{\bsm}{\begin{smallmatrix}}
\newcommand{\esm}{\end{smallmatrix}}

\newcommand{\ddiffp}[2]{\frac{{\p}^2 #1}{{\p #2}^2}}

\newcommand{\be}{\begin{equation}}
\newcommand{\ee}{\end{equation}}

\renewcommand{\d}{{\mathrm d}}

\newcommand{\R}{\mathbb{R}}

\newcommand{\braket}[2]{\langle #1 | #2 \rangle}

\newcommand{\MSC}[1]{\\[1em]{\footnotesize MSC (2000): #1}\\ \vskip-3.2em\phantom{}}

\begin{document}

\preprint{AEI-2013-001}

\title{Solutions of the Klein-Gordon equation in an infinite square-well potential with a moving wall}

\author{Michael Koehn}
\email[]{michael.koehn@aei.mpg.de}
\affiliation{Max-Planck-Institut f\"ur Gravitationsphysik, Albert-Einstein-Institut, 
Am M\"uhlenberg 1, 14476 Potsdam, Germany, EU
}

\begin{abstract}
Employing a transformation to hyperbolic space, we derive in a simple way exact solutions for the Klein-Gordon equation in an infinite square-well potential with one boundary moving at constant velocity, for the massless as well as for the massive case.
\MSC{35R37, 81Q05, 81Q35, 35Q40, 35Q41}
\end{abstract}

\maketitle

{\bf~Introduction.}~-- The {\it non-relativistic} system of a one-dimensional infinite square well with a massive particle evolving according to the Schr\"odinger equation is one of the most elementary quantum mechanical systems, and it often serves as an approximation to more complex physical systems. If, however, the potential walls are allowed to move, as originally in the Fermi-Ulam model for the acceleration of cosmic rays \cite{Fer49,Ula61}, the situation is much more complicated: if one does not choose to rely on an adiabatic approximation, then the system is not separable anymore. Results concerning exact solutions exist only sparsely and have attracted a large amount of attention. For the special case of a non-relativistic system with a wall moving at constant velocity, such exact solutions have been obtained first in \cite{DR69}, see \cite{MD91,DMP95,GMNN09,SLMSA09,Mou12} for generalizations.

The {\it relativistic} moving-wall system on the other hand is a much less common object of study (though it has interesting applications in graphene billiards and cosmological billiards), and there appear interesting subtleties. This article concerns the one-dimensional Klein-Gordon (KG) particle in an infinite square well with a boundary which is moving outward at constant velocity $\nu$. We find an infinite set of exact solutions which do not rely on an adiabatic approximation on top of the square-well approximation with definite position and momentum configuration. We thereby generalize the solutions presented in the appendix of \cite{Rya72}, which are valid only for a special case and which are stated without specifying any method on how to obtain them. In contrast, we use a transformation to hyperbolic space which provides in a simple and new manner a set of general solutions for the massless as well as for the massive case.

{\bf~Exact~solution.}~--
This article concerns the initial/boundary value problem
\begin{align}\label{KGm}
\begin{cases}
\ddiffp{}{t}\Psi(t, x)=\ddiffp{}{x}\Psi (t,x)-m^2\Psi(t,x) &\quad \text{in }\F \\
\Psi\vert_{\p\F}=0&\quad\text{on }\p\F\\
\Psi(t_0,x)= f(x), \  (\p_t\Psi)(t_0,x)=g(x). &
\end{cases}
\end{align}
\begin{figure}
\includegraphics{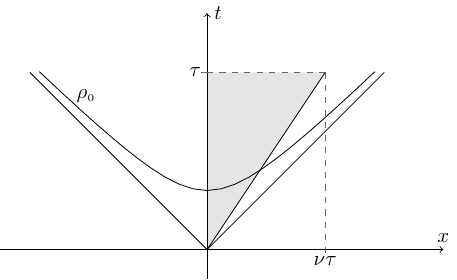}
\caption{The infinite square well with moving walls as a shaded wedge inside the forward lightcone, intersected by a spacelike hyperboloid of fixed $\rho=\rho_0$.}\label{fig:kegel}
\end{figure}
The infinite square well is specified as the domain $\F= [ 0 , L(t) ]$. It is given in terms of the length function $L(t)=L_0+\nu(t-t_0)$, where $\nu$ denotes the speed of the receding wall with $0<\nu<1$. $\p\F$ denotes the boundary of $\F$, and $t, x\in\R$. We have rescaled the speed of light and Planck's constant to $c=1=\hbar$, and we will treat the massless case $m=0$ first. Since \eqref{KGm} is of second order in time, the transformation
\be
x \mapsto x'=\frac{x}{L(t)}
\ee
which is usually applied in the non-relativistic scenario (see e.g.~\cite{MD91,DMP95}) is of not much help in obtaining analytic solutions for the relativistic one, although indeed implying motionless walls for the transformed system.

However, by slicing the forward lightcone of two-dimensional Minkowski space in terms of hyperboloids and transforming to hyperbolic coordinates (cf.~Fig.~\ref{fig:kegel}), one obtains a separable infinite square well system with static boundaries, where the hyperbolic radial coordinate $\rho$ serves as the new time-like coordinate. Explicitly, the transformation is given by
\be
t=\rho\gamma_t\qquad x=\rho\gamma_x
\ee
subject to the constraint
\be
(\gamma_t)^2-(\gamma_x)^2=1
\ee
which implies that $\rho^2=t^2-x^2$.

\begin{figure}
\centering
\includegraphics{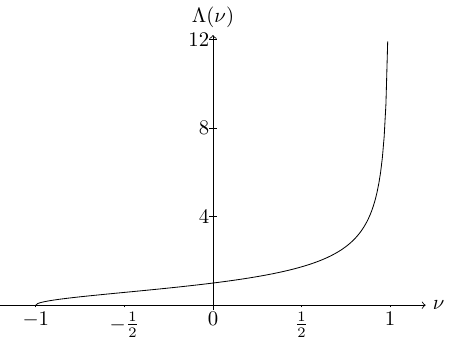}
\caption{The position $\Lambda$ of the boundary in hyperbolic space depends only on the speed $\nu$ of the moving wall.}\label{fig:LambdaPlot}
\end{figure}

The coordinates $\gamma_t$ and $\gamma_x$ are constrained to the one-dimensional hyperboloid, such that one of the two coordinates is redundant and we may parametrize the hyperboloid as the one-dimensional analogue of the hyperbolic upper half-plane through the transformations
\be\label{UHP_1d}
\gamma_t =\frac{v^2+1}{2v}\qquad\gamma_x=\frac{v^2-1}{2v} \qquad v=\gamma_t+\gamma_x\ .
\ee
One can easily find a full set of exact solutions in these coordinates by requiring as an intermediate step that $L_0=\nu t_0$. In order to obtain solutions for general parameters $(L_0,t_0)$, one may afterwards freely shift the tip of the lightcone. The position of the right boundary does not depend on the time-like coordinate $\rho$, but merely on the constant velocity $\nu$. Employing \eqref{UHP_1d}, the position of the right boundary in hyperbolic space can be determined as (cf.~FIG.~\ref{fig:LambdaPlot})
\be\label{Lambdafunction}
\Lambda(\nu)=\sqrt{\frac{1+\nu}{1-\nu}} \ ,
\ee
\begin{figure}
\centering
\includegraphics[width=\columnwidth]{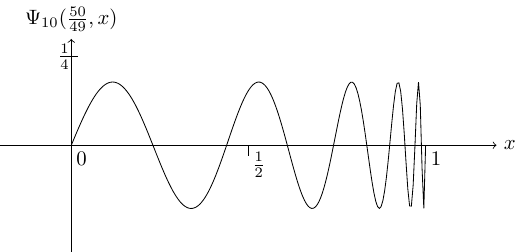}
\caption{Exemplary plot of the wavefunction $\Psi_{10}(\frac{50}{49},x)$ with speed $\nu=\frac{49}{50}$ of the moving wall at time $t=\frac{50}{49}$.}\label{fig:soln}
\end{figure}
while the transformed massless KG equation reads
\be\label{KGhyp}
\rho\p_\rho\left(\rho\p_\rho\Psi(\rho,v)\right)=v\p_v\left(v\p_v\Psi(\rho,v)\right) \ .
\ee
Due to the separability of the system in hyperbolic coordinates, the general solution is given by
\be
\Psi(\rho,v)=\psi_1(\rho v)+\psi_2(\rho/v) \ ,
\ee
where $\rho v=t+x$ and $\rho/v=t-x$, and we can easily choose a solution that matches the Dirichlet boundary conditions, e.g.
\be\label{kg1dsol_hyp}
\Psi_n(\rho,v)=\frac1{\sqrt{n\pi}}\sin\left(k_n\ln(v)\right)\exp(-\I k_n\ln(\rho))
\ee
with
\be
k_n=\frac{n\pi}{\ln(\Lambda(\nu))} \ .
\ee
For the massive case corresponding to $\p_t^2\Phi=\p_x^2\Phi-m^2\Phi$ with $m\neq 0$, the above procedure implies the solutions
\be
\Phi_n(\rho,v)=C \sin\left(k_n\ln(v)\right)\left[J_{\I k_n}(m\rho)+Y_{\I k_n}(m\rho)\right] \ ,
\ee
where $J_{\I k}$ and $Y_{\I k}$ are the Bessel functions of imaginary order $\I k$. Of course we could have also chosen real solutions to the real KG equation. We remark that in a general number of dimensions, having a definite direction of wavepacket propagation is related to complexity of the wavefunction (cf.~also \cite{Koe12}).

We have chosen the integration constants of the solutions \eqref{kg1dsol_hyp} such that they are normalized to $1$ with respect to the KG-like scalar product
\be\label{hypscalar}
\left(\psi_1,\psi_2\right)= \I \rho\int\d \mathrm{vol}\psi_1^*\nquad\stackrel{\leftrightarrow}{\p_\rho}\nquad\psi_2 \ ,
\ee
where $\d\mathrm{vol}=\frac{\d v}{v}$ denotes the volume element on the one-dimensional hyperbolic upper half-plane, and where $\psi\nquad\stackrel{\leftrightarrow}{\p_\rho}\nquad\phi\equiv\psi\p_\rho\phi-\phi\p_\rho\psi$.
The solutions \eqref{kg1dsol_hyp} can be expressed in flat coordinates as
\begin{align}\label{kg1d_movsolns}
\Psi_n(t,x) = \frac1{\sqrt{n\pi}}\sin\left[\frac{k_n}2\ln\left(\textstyle\frac{t'+x}{t'-x}\right)\right]\times\notag\\
\times\exp\left[-\I \frac{k_n}2\ln\left(t'^2-x^2\right)\right]
\end{align}
with $t'(t)\equiv t-t_0+\frac{L_0}{\nu}$, $0<\nu<1$. FIG.~\ref{fig:soln} displays an exemplary plot of one specific such function. We note that the solutions \eqref{kg1d_movsolns} are normalized to $1$ with respect to the standard KG-invariant scalar product,
\be\label{KG_scalar}
\braket{\psi_1}{\psi_2} = \I\int \d x\psi_1^* \nquad \stackrel{\leftrightarrow}{\p_\loOm} \nquad \psi_2 \ .
\ee
The respective norms in hyperbolic and in flat space are preserved,
\begin{figure*}[t]
\subfigure[Before reflection]{\includegraphics[width=0.24\textwidth]{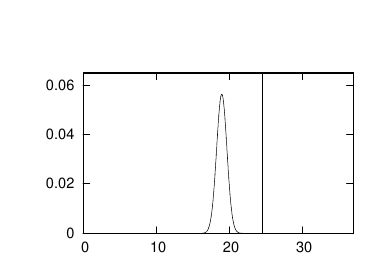}}
\subfigure[During reflection]{\includegraphics[width=0.24\textwidth]{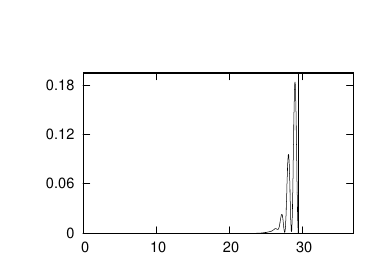}}
\subfigure[Redshifted wavepacket]{\includegraphics[width=0.24\textwidth]{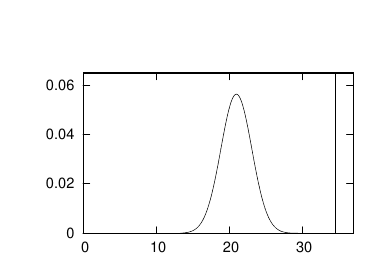}}
\subfigure[Late-time behavior (scale changed)]{\includegraphics[width=0.24\textwidth]{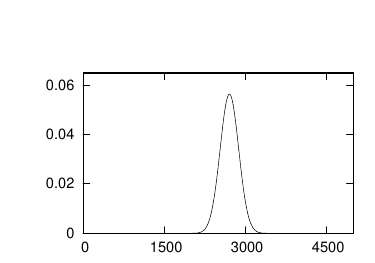}}
\caption{Snapshots of $|\psi(t,x)|^2$ for a one-dimensional Gaussian wavepacket evolving with respect to \eqref{KGm} and redshifted upon reflection off a wall moving in flat space. The wall is moving at $\nu=\frac12$ and is represented by the vertical bar in (a)-(c). The scaling in (d) differs from (a)-(c).}\label{fig:1dflat}
\end{figure*}
\begin{figure*}[t]
\subfigure[Before reflection]{\includegraphics[width=0.24\textwidth]{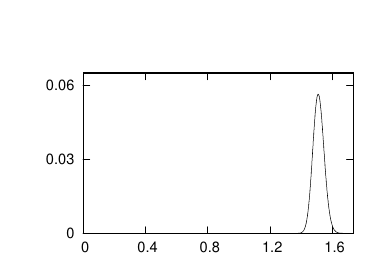}}
\subfigure[During reflection]{\includegraphics[width=0.24\textwidth]{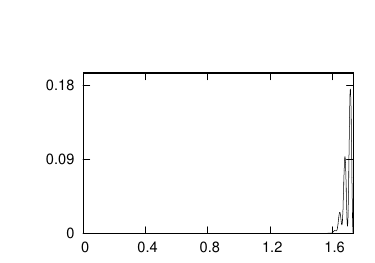}}
\subfigure[Reflected wavepacket]{\includegraphics[width=0.24\textwidth]{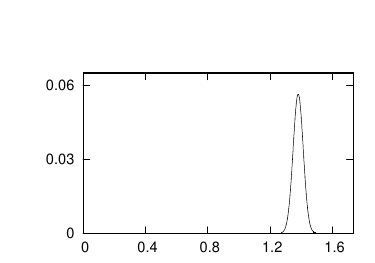}}
\subfigure[Late-time behavior]{\includegraphics[width=0.24\textwidth]{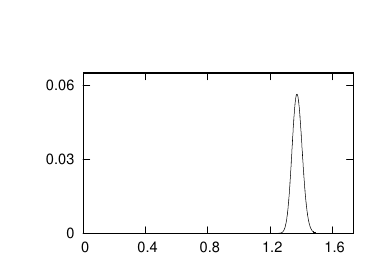}}
\caption{Snapshots of the same $|\psi(t,x)|^2$ for a one-dimensional Gaussian wavepacket, but this time with $t$ and $x$ expressed through the hyperbolic coordinates $\rho$ and $v$ evolving with respect to \eqref{KGhyp} and reflecting off an infinite square well with right boundary at $\Lambda(\frac12)=\sqrt3$ according to \eqref{Lambdafunction}.}\label{fig:1dhyperb}
\end{figure*}and we can directly show for \eqref{kg1d_movsolns} that
\be
\braket{\Psi_n}{\Psi_m} = \delta_{nm}\ .
\ee

{\bf Numerical~investigations.}~--
We now investigate the properties of a relativistic quantum wavepacket evolving according to \eqref{KGhyp}, i.e.~$m=0$. For this purpose, we compose out of positive frequencies $\omega=|p|$ a one-dimensional Gaussian wavepacket
\be\label{1pac}
\Psi(t,x) = A \int_{-\infty}^\infty\d p \e^{-\frac{c^2(p-p_0)^2}2+\I(px-\omega t)}
\ee
which is normalized to $1$ with respect to \eqref{KG_scalar} if
\be
A = \frac{c}{\sqrt\pi}\left( \e^{-c^2p_0^2}+\sqrt\pi p_0c\ \erf(p_0c) \right)^{-\frac12}\ ,
\ee
where $\erf$ denotes the error function. The norm of the wavepacket \eqref{1pac} with respect to \eqref{KG_scalar} (approximately $1$ if $c$ is chosen small enough) is preserved during its evolution in the moving-wall system. However, its absolute width with respect to the $x$-coordinate grows with each reflection off the moving wall due to successive {\it redshifts}. The redshift upon reflection of a one-dimensional wave off a wall which is moving away at a constant speed is given by
\be
1+z = \frac{f_\mathrm{ref}}{f_\mathrm{inc}} = \gamma^2 (1+\nu)^2 = \frac{1+\nu}{1-\nu} \equiv \Lambda(\nu)^2\ ,
\ee
with $\gamma=(1-\nu^2)^{-\frac12}$, e.g.~$1+z=3$ for $\nu=\frac12$. The corresponding reflection of a classical massless particle in one dimension off a moving wall in terms of the relativistic Hamiltonian $H=\sqrt{{\pi_x}^2}$, where $\pi_x$ denotes the momentum variable conjugate to $x$, implies the relation
\be
1+z = \frac{H_\mathrm{ref}}{H_\mathrm{inc}}
\ee
of the energy $H_\mathrm{inc}$ of a photon before a bounce from a moving mirror to its energy $H_\mathrm{ref}$ afterwards.
The expectation value $\langle \hat H \rangle$, which we call the energy expectation value in the following, may serve as a measure for the redshift. The square root in the Hamiltonian can be avoided adopting the two-component notation \cite{FV58} for \eqref{KGm} (especially helpful in higher-dimensional cases),
\be
\psi = \phi+\chi \quad\I\p_\loOm\psi=\phi-\chi\ ,
\ee
which we adapt to the massless case to obtain
\be\label{Hsq}
\hat{H} = -\frac{\sigma_3+\I\sigma_2}2\Delta+\frac{\sigma_3-\I\sigma_2}2 \ ,
\ee
where $\sigma_i$ denote the Pauli matrices and $\Delta=\ddiffp{}{x}$. With the two-component vector
\be
\Psi = \left(\bsm \phi \\ \chi \esm\right) \ ,
\ee
the inner product \eqref{KG_scalar} is expressible as
\be\label{KGvector}
\langle \Psi_1 | \Psi_2 \rangle = 2 \int \d x \Psi_1^\dagger \sigma_3 \Psi_2
\ee
and the energy expectation value as
\be
\langle\hat H \rangle = 2\int \d x \Psi^\dagger \sigma_3 \hat{H}\Psi \ .
\ee
Furthermore, \eqref{Hsq} implies the expected expression
\be
\langle\hat{H}^2 \rangle=2\int\d x \Psi^\dagger\sigma_3\hat{H}^2\Psi= -\I\int \d x \psi^* \nquad\stackrel{\leftrightarrow}{\p_\loOm}\nquad(\Delta\psi) \ .
\ee
FIG.~\ref{fig:1dflat} and FIG.~\ref{fig:1dhyperb} illustrate the reflection of the wavepacket off the wall, obtained as the numerical solution of the massless KG equation, on the one hand in flat space with moving boundary conditions acording to \eqref{KGm}, and on the other hand the corresponding evolution in hyperbolic space with static boundary conditions according to \eqref{KGhyp}. In flat space, the wavepacket gets redshifted and loses energy upon reflection off the moving wall, while its KG norm \eqref{KG_scalar} is preserved. We remark that unlike the Hamiltonian and momentum operators, the position operator generally mixes positive and negative frequency components of the wavepacket in relativistic quantum mechanics \cite{FV58}, and we refer the reader to the investigations in \cite{NW49,Wig62} for details on the localization of relativistic particles.

{\bf Summary~and~conclusions.}~--
Exact solutions of the Schr\"odinger and Klein-Gordon equations in a domain with time-dependent boundaries are notoriously difficult to obtain. Although solutions for the Schr\"odinger equation and also for the d'Alembert equation in a domain with special types of boundary movements are known, this was not the case for the Klein-Gordon equation. With this letter, we contribute an infinite set of orthogonal exact solutions to the one-dimensional Klein-Gordon equation in an infinite square-well with one wall moving at a constant velocity. These solutions are obtained employing a simple transformation to hyperbolic space. We furthermore investigated numerically the properties of a massless relativistic wavepacket bouncing off the moving walls in flat and off the static walls in hyperbolic space, and in the former case observed the expected redshift. Although the scope of this article was intended to be limited to a domain with one wall moving at constant speed, it would nevertheless be of interest to generalize 
these results to domains with more arbitrarily moving walls.

\begin{acknowledgments}
The author gratefully acknowledges the support of the European Research Council via the Starting Grant numbered 256994.
\end{acknowledgments}

\bibliographystyle{apsrev}
\bibliography{../../../bib/master}

\end{document}